\input amstex.tex

\define\gm{\bold g}

\define\<#1,#2>{\langle #1,#2\rangle}
\define\tr{\text{tr}}
\define\id{I} 
\define\TR{\text{TR}} 
\define\Res{\text{Res}}
\define\QED{\hfill$\square$}
\define\omprime{\omega'} 
\define\omloc{\omega^{loc}}

\documentstyle{amsppt}
\advance\vsize -2cm
\magnification=1200
\hfuzz=30pt
\NoBlackBoxes

\topmatter
\title Anomalies and Schwinger terms in NCG field theory models
\endtitle
\author E. Langmann, J. Mickelsson, and S. Rydh \endauthor
\affil  Theoretical Physics, Royal Institute of Technology, SE-10044,
Stockholm, Sweden \endaffil 
\date March 1, 2001 \enddate
\endtopmatter
\document
ABSTRACT  We study the quantization of chiral fermions coupled to 
generalized Dirac operators arising in NCG Yang-Mills theory. 
The cocycles describing chiral symmetry breaking are calculated. In
particular, we introduce a generalized locality principle for the
cocycles. Local cocycles are by definition expressions which can be 
written as generalized traces of operator commutators. In the case of
pseudodifferential operators, these traces lead in fact to integrals 
of ordinary local de Rham forms. As an application of the general
ideas we discuss the case of noncommutative tori. 
We also develop a gerbe theoretic 
approach to the chiral anomaly in hamiltonian quantization of NCG 
field theory. 

\vskip 0.3in

1. INTRODUCTION  

\vskip 0.3in
The nonabelian chiral anomaly appears in two different (but related) 
ways. First, it modifies the structure of current algebra.
Historically, this was the way quantum mechanical symmetry breaking
was first observed, [1], but it took some time before 
a clear mathematical formulation of this phenomenon was obtained, 
[2].  On the other hand,  the anomaly appears in 
the path integral formulation as a lack of gauge invariance of the 
effective action, [3].  In either way, the anomaly can be 
thought of as an element of an appropriate group (or Lie algebra) 
cohomology group. 

The nonabelian gauge anomaly arises from a left-right nonsymmetric 
coupling of vector potentials to a Dirac fermion field.  This process 
can be generalized in a straight-forward way in noncommutative
geometry models. Instead of a standard Dirac operator one considers 
self-adjoint (invertible) unbounded operators  $D$ with the property that $1/|D|^p$ 
is 'almost' trace-class for a given real number $p\geq 1.$  'Almost' 
means that the spectral integral 
$\int_{0}^{\Lambda}  (|\lambda|^p +\epsilon)^{-1} d\mu(\lambda)$ 
is at most logarithmically divergent at 
$\Lambda\to\infty,$ for any positive number $\epsilon.$
After fixing such an operator we study
generalized Dirac operators $D_A= D +A$ where $A$ is any hermitean
bounded perturbation.     
    
We stress that our considerations are very general:  The family of 
Dirac operators could for example arise from a coupling of vector 
potentials through a star product (generalized Moyal brackets).  All 
what is needed is that the $L_p$ estimate mentioned above is valid. 
This has been proven to hold in a class of star product quantizations 
[4] defined by a constant antisymmetric $\theta$ 
matrix. We shall study some consequences of this in section 7.

Our starting point for a  construction of a NCG field theory model is 
a triple $(D, \int, \Cal B),$ where $D$ is the Dirac operator acting 
in a Hilbert space $H$, $\Cal B$ is an associative algebra of
operators in $H$ such that $[D,b]$ is bounded for all $b\in\Cal B.$ 
In addition,  there is an 'integration',  or a generalized trace map, from 
operators ('p-forms') of the type $\omega=b_0[\epsilon,b_1][\epsilon,b_2]\dots
[\epsilon,b_p],$ or alternatively $|D|^{-p}b_0[D,b_1]\dots [D,b_p],$ 
to complex numbers, [5], see also the discussion around Definition
10.8 in [24] for more details.   Here
$\epsilon=D/|D|$ is the sign of the Dirac operator. In some cases (but
not always) one can prove an equality between the integrals of the two alternative 
expressions, [5].  The generalized vector potentials are then 
linear combinations of 1-forms $A=b_0[D, b_1],$ or sometimes
$a=b_0[\epsilon,b_1].$ 
  
We study the BRST double complex based on de Rham forms on the space 
of vector potentials $A$ with values in the space of the above
operator  p-forms  $\omega,$ with $p=0,1,2,\dots.$ The NCG BSRT
complex has been previously studied from different points of view in 
[6] and  more recently in [7].  

One of the central results in the present paper concerns the question 
of locality of the BRST cocycles. In abstract NCG models it is not 
a priori clear what this could mean because in the operator approach 
there is in general no space-time manifold to define local fields. 
However, in the ordinary space-time geometric setup many of the basic 
quantities can be written as 'traces' of commutators of
pseudodifferential operators. For example, this holds for nonabelian
Schwinger terms [8] and the gauge anomaly in the path integral 
formulation [9]. It turns out that these traces are in fact 
integrals of local differential forms.  We adopt this as our starting 
point: We prove that in the NCG field theory models there is a 'local' 
anomaly formula expressing the BRST cocycles as traces of commutators 
of nontrace-class operators.  

The second main result concerns the description of the hamiltonian 
anomaly in gerbe theoretic terms in which the basic object is a
3-cohomology class, the Dixmier-Douady class.  The results are 
largely generalizations of the corresponding considerations for the standard 
coupling of local vector potentials to chiral fermions [10]. 
However, there is one important technicality which makes  a 
difference between the 'classical' and  the NCG case. In the classical
case the Dixmier-Douady class defines a de Rham form on the space of
gauge orbits $\Cal A/\Cal G.$ This is a smooth manifold when one
restricts to the case of so-called based gauge transformations $\Cal
G_0.$   In the NCG models the concept of based gauge transformations is
not well-defined and the space of gauge orbits has singularities. 

In order to avoid these singularities we realize that space of NCG
gauge potentials for the Schatten index $p$ as loops in gauge
potentials for index $p-1.$ Correspondingly, the group of gauge
transformations for index $p$ is the group of smooth loops in the 
gauge group for the case of index $p-1$ which allows us to define 
based gauge transformations in the usual way as loops pass the neutral 
element at a fixed value of the argument.  This leads then to a
construction of a gerbe and its Dixmier-Douady class as a de Rham form 
in a standard way. 

We want to thank G\"oran Lindblad for drawing to our attention the
reference [23].

\vskip 0.3in

2. THE GENERAL SETUP FOR NCG DESCENT EQUATIONS AND ANOMALIES 

\vskip 0.3in
Let $D_0$ be an unbounded selfadjoint operator in a complex Hilbert 
space $H$ such that $|D_0|^{-1} \in L_{p+},$ that is, $|D_0|^{-p}\in  
L_{1+}$ for some $p\geq 1.$ Here $L_{1+}$ is the Dixmier ideal in the 
algebra of bounded operators in $H.$ A positive operator $T$ is in
$L_{1+}$ if it is compact and 
$$\frac{1}{\log N} \sum_{k=1}^N \lambda_k$$ 
has a finite limit, where $\lambda_1\geq \lambda_2\geq\dots$ are the
eigenvalues of $T.$  
We also assume for simplicity that
$D_0$ is invertible.  However, the following discussion can be 
easily generalized to the case when $D_0$ is a finite rank
perturbation of an invertible operator.   

We shall work with bounded perturbations of the 'free Dirac operator'
$D_0,$ denote $D_A = D_0 +A$ where $A$ is a bounded selfadjoint
operator in $H$ such that $[|D_0|,A]$ is bounded. 
We shall denote $F_A$ a smoothed sign  operator associated to $D_A.$ 
The technical complication is that the  map $A\mapsto D_A/|D_A|$ is not
continuous when $D_A$ has zero modes.  Instead, we can take a smooth
function $f:\Bbb R \to \Bbb R_+$ such  that $f(x)-|x|$ approaches zero 
faster than any power of $x$ as $|x|\mapsto \infty$ and $f(x)\geq m$
for some positive constant, and we define $F_A= D_A/f(D_A).$ For
example, take $f(x)= +\sqrt{x^2 + e^{-x^2} }.$  Then $A\mapsto F_A$ is 
norm continuous. If $D_A$ is the classical Dirac operator associated
to a vector potential $A$ on a compact manifold then the difference 
$D_A/|D_A| -F_A$ is an infinitely smoothing pseudodifferential
operator and in particular a trace class operator.  

The sign  operator $D_A/|D_A|$ defines a Fock space representation of
the canonical anticommutation relations algebra CAR.  The generators
in the CAR algebra are denoted by $a(v), a^*(v),$ where $v\in H.$ The 
algebra is defined by the basic relations 
$$\align a(u)a(v) +a(v)a(u)& =0= a^*(u)a^*(v)+a^*(v)a^*(u) \\ 
a^*(u) a(v) + a(v) a^*(u) & = <v,u>,\tag2.1\endalign$$ 
where the Hilbert space inner product $<.,.>$ is antilinear in the
first argument.  The Dirac representation is then fixed by the
requirement that there is a \it vacuum vector \rm $|A>\in\Cal F_A$
such that 
$$a^*(u)|A> =0= a(v) |A> \text{ for $u\in H_-(A)$ and $v\in H_+(A)$ }
\tag2.2$$
where $H=H_+(A) \oplus H_-(A)$ is the polarization to the spectral
subspaces $D_A\geq 0,$ $D_A <0,$ respectively. 

Because of the potential zero modes of the Dirac operator the Dirac 
vacuum cannot in general be defined as a continuous function of $A.$ 
Instead, we must be content with a choice of a polarization
$H=W(A)\oplus W(A)^{\perp} $ such that $W(A) \sim H_+(A),$ where the
equivalence is defined by the condition that the projection operators 
to the given subspaces differ only by Hilbert-Schmidt operators. 
The Hilbert-Schmidt condition comes from the requirement that the CAR 
representations defined by the two polarizations should be equivalent;
for a review on CAR representations see 
[11].  Let $\epsilon_{W}= \pi_{W} -\pi_{W^{\perp}}$ the
grading operator defined by the orthogonal projection $\pi_W:H\to W.$ 
Then $\epsilon_{W(A)} - \epsilon_{H_+(A)}$ is Hilbert-Schmidt and
$\epsilon_{H_+(A)} -F_A$ is trace class and so $F_A$ is an
approximation of $\epsilon_{W(A)}$ modulo Hilbert-Schmidt operators. 
The advantage of working with $F_A$ is that it is easier to produce 
explicit formulas (as we saw above), compared with the grading
operators $\epsilon_{W(A)}.$     

As an operator function $D_A/f(D_A)$ of $D_A,$ the operator $F_A$
satisfies $g^{-1}F_A g = F_{A^g},$ with $A^g= g^{-1} A
g+g^{-1}[D_0,g]$ for a unitary transformation $g$ such that $[D_0,g]$ 
is bounded.  Denote by $\Cal B$ the algebra of bounded
operators $b$ in $H$ such that $[D_0,b]$ and $[|D_0|,b]$ are
bounded. Then all the operators $A=b_0[D_0,b_1],$ for $b_i\in \Cal B,$
satisfy the condition $[|D_0|,A]$ is bounded.  We denote by
$U_{p+}$ the group of unitary elements in $\Cal B.$ Any element $g\in
U_{p+}$ satisfies $[\epsilon, g] \in L_{p+}$ where $\epsilon =
D_0/|D_0|.$

\proclaim{Lemma 2.3}  $F_A-\epsilon \in L_{p+}$ for any bounded operator
$A$ such that $[|D_0|,A]$ is bounded. \endproclaim

\demo{Proof} 1) We first prove that $|D_0+A|-|D_0|$ is bounded.
We have assumed that $D_0$ is invertible, so $|D_0|\geq
\mu$ for some positive constant $\mu.$ 
We use the norm estimate $|| |X| -|Y| || \leq || |X^* X -
Y^*Y|^{1/2} ||,$ see for example [23, Section X.2].  Set $G=D_0A+ AD_0$
and split $G= G_+ +G_-$ where $G_+$ commutes with $\epsilon$ and $G_-$
anticommutes with $\epsilon.$ The linear equation 
$$ G = |D_0| Z + Z |D_0|$$ 
can be solved as $Z= Z_+ +Z_-,$ with $Z_+ = A_+ \epsilon$ and 
$$Z_- = \int_0^{\infty} e^{-t|D_0|} [|D_0|, \epsilon A_-] e^{-t|D_0|}
dt,$$  
and thus $Z$ is bounded, since  $||e^{-t|D_0|}|| \leq e^{-t\mu}$ and
since $[|D_0|,\epsilon A_-]=\epsilon[|D_0|,A_-]$ is bounded by
assumption. Setting $X= D_0 A$ and $Y=|D_0| + Z$ in the above norm 
estimate we observe that $|D_0+A| - ||D_0| + Z|$ is bounded. In the
case when $||Z|| \leq \mu$ the operator $|D_0| + Z$ is positive and so 
$||D_0| +Z| = |D_0| +Z$ and it follows that
$|D_0+A|-|D_0|=(|D_0+A|-||D_0|+Z|) +(||D_0| +Z| -|D_0|)$ is bounded. 
In the general case, replace $D_0\mapsto D'=D_0 +i\alpha$ for a real number
$\alpha.$ Clearly $|D_+i\alpha|-|D_0|$ and $|D_0+A+i\alpha|-|D_0+A|$
are bounded, so the question whether $|D_0+A|-|D_0|$ is bounded is
equivalent to whether $|D'+A| -|D'|$ is bounded. Now $|D'| \geq
|\alpha|$ and so we have $||Z|| \geq |D'| $ when we choose $\alpha$
large enough and the proof reduces to the special case above. 

2) We may assume without essential restriction that  $D_0 +A$ is 
invertible so we can select $f(x)=|x|$ and 
$$\align F_A -\epsilon&= (D_0+A)^{-1}|D_0+A| - D_0^{-1}|D_0| = 
(1+ D_0^{-1} A)^{-1} D_0^{-1} |D_0+A| -D_0^{-1}|D_0|\\
&=  \{(1+D_0^{-1} A)^{-1} -1\} \epsilon + (1+D_0^{-1} A)^{-1} D_0^{-1}
T,\endalign $$ 
for some bounded operator $T.$ The second term on the right is then in
$L_{p+}$ since $D_0^{-1} \in L_{p+}$ and the first term is of the form 
$((1+S)^{-1}-1)\epsilon$ with $S\in L_{p+}$ and thus it belongs also
to $L_{p+}.$    {\QED}\enddemo
   
The following discussion is based on a parameterization of the
fermionic Fock space by the operator $a=F_A -\epsilon.$ Thus the Dirac
vacuum for the vector potential $A$ is given by any grading operator
$\epsilon_a$ 
which differs from $\epsilon +a$ by a Hilbert-Schmidt
operator. Geometrically, this leads to the construction of a smooth
infinite-dimensional vector bundle $\Cal F$ over the parameter space
$\Cal A$ of generalized vector potentials. The fiber $\Cal F_a$ is the 
Fock space defined by the grading operator $\epsilon_a.$  The
quantized Dirac operator $\hat D_A$ acts as an unbounded positive operator in the 
fiber.  

The group $U_{p+}$ acts in the base space of the bundle. The problem
arises whether the action can be lifted to the total space such that 
$g^{-1}\hat D_A g = \hat D_{A^g}.$ In case of smooth vector
potentials, $A= \gamma^k A_k,$ and massless Dirac operator it is known
that the answer is negative. Instead, there is an extension $\hat \Cal
G$ of the group of gauge transformations $\Cal G$ which acts in the total
space. The obstruction to the $\Cal G$ action is the extension term in
the commutation relations of the Lie algebra $Lie(\Cal G).$ This
extension (Schwinger term) is a 2-cocycle $c_{n,2}$ of $Lie(\Cal G)$ in the
module of complex functions of the variable $A.$ In the case of one
space dimension, $n=1,$ the cocycle does not depend on $A$ and one may 
restrict to the module of constant functions. 

There is an equivalent alternative way to view the obstruction. The
standard construction in canonical quantization leads naturally to a
bundle $P\Cal F$ of \it projective Fock spaces \rm which do not depend
on any choices of the grading operator $\epsilon_a.$ The existence of
a Fock bundle which gives $P\Cal F$ as its projectivization is then 
related to triviality of a \it Dixmier-Douady \rm class in $H^3(\Cal
A/\Cal G, \Bbb Z),$ [10].  

The main content of the present article is to explain how the local
formulas for gauge anomalies, Schwinger terms, the Dixmier-Douady
class, and all the cocycles related to these through the standard BRST 
descent equations extend to certain cocycles on the space of bounded 
operators $A$ and the Lie algebra $Lie(U_{p+})$ such that a
restriction to the classical case gives standard local formulas. A
central ingredient is to use (noncyclic) extensions of the trace functional to 
nontrace-class operators. 
   
The 'infinitesimal version' of the gauge transformation $A\mapsto A^g$
in terms of the parameter $a$ is
$$\delta_X a = [a,X] + [\epsilon, X] \text{ for $X\in \bold{u}_p=Lie(U_{p+})$ }.\tag2.4 $$  
 
Let us recall the basic definitions in NCG differential calculus for
Fredholm modules
[5]. The differentials of order $n$ are linear combinations of operators 
of the type $b_0[\epsilon, b_1]\dots [\epsilon, b_n]$ where $b_i \in
\Cal B.$ One denotes $db= [\epsilon, b]$ for $b\in \Cal B.$ If
$\phi\in\Omega^n$ is a differential of order $n$ then $d\phi =
\epsilon \phi +(-1)^{n+1}\phi \epsilon.$ This gives a map $d:\Omega^n
\to \Omega^{n+1}$ with $d^2=0.$  The cohomology of this complex is
trivial.  

The coboundary operator associated to infinitesimal gauge
transformations is denoted by $\delta.$ We work with cochains of 
order $k,$ $\tau\in \Omega_k,$ consisting of functions $\tau(a;
X_1,\dots, X_k)$ of $a\in \Omega^1$ and of Lie algebra elements 
$X_i\in \bold{u}_p,$ linear in each $X_i$ and totally antisymmetric 
in the arguments $X_i.$ The standard Lie algebra coboundary operator is defined by 

$$\align (\delta\tau_n)(a; X_1,\dots, X_{n+1})& = \sum_i (-1)^{i-1}\delta_{X_i} \tau(a;
X_1, ..., \hat X_i, ...,X_{n+1}) + \\
& \sum_{i<j}  (-1)^{i+j} \tau(a;[X_i,
X_j],.., \hat X_i,...,\hat X_j,...,X_{n+1}),\tag2.5\endalign $$ 
where the hat means that the corresponding argument is deleted and
$\delta_{X_i}$ is the Lie derivative acting on functions of $a,$ the
action on the argument being given by (2.4). We remind that the
multilinear forms $\tau$ on a Lie algebra can be interpreted as 
left invariant differential forms on the corresponding Lie group (and
vice versa) through the standard identification of a Lie algebra as 
left invariant vector fields.  

We shall also work with a $(d,\delta)$ double complex consisting of
$\delta$ forms taking values in the $d$ complex $\Omega^*$, using
the standard BRST sign convention. The BRST ghost $v$  can be 
interpreted  as the Maurer-Cartan 1-form on the gauge group, that is, 
at the identity element it is the
the  tautological 1-form sending a Lie algebra element onto itself,
$v= g^{-1}\delta g.$ The sign conventions are encoded into the
algebraic rules 
$$
   \align d^2 &= \delta^2= \delta d + d \delta = 0 \\
   \delta(v)& = -v^2 = \delta v +v \delta \\
   \delta(a) &= -[a,v]_+ -v^2 = \delta a + a\delta \\
    d(v)& = [\epsilon,v]_+ = d v +vd\\
    d(a)&  = [\epsilon,a]_+ =da +a d.\tag2.6\endalign  $$
We denote $[a,b]_+=ab+ba.$

Here a remark on notation is in order. We want to treat at the same time
both the even and odd Fredholm modules (related to odd/even $k.$)
To obtain the correct signs in (2.6), we need to make a
reinterpretation of $x=v,\delta,\epsilon$ and $a$. 

\it Even Fredholm module. \rm In this case we have, 
by definition, a hermitean operator $\Gamma$ in $H$ with $\Gamma^2=1$
which anticommutes with $\epsilon$ and $a$, and the correct signs
are accounted for if we interpret the ghost as $\Gamma v$.  To be
precise, one should distinguish between $v,\delta,\epsilon$ and $a$ 
and
$$
\align
s(v) &= \Gamma v, \quad
s(\delta) = \Gamma\delta  .\\
s(\epsilon) &= \epsilon, \quad  
s(a) = a . \tag2.7
\endalign
$$

\it Odd Fredholm module. \rm We do not
have a $\Gamma$ at hand but we introduce it by doubling the original
Hilbert space, $H\mapsto H\otimes \Bbb C^2,$ and introducing
the usual Pauli sigma matrices $\sigma_3$ and $\sigma_1$ acting on the 
second factor. We can then define
$$
\align
s(v) &= v\otimes \sigma_1, \quad
s(\delta) = \delta\otimes \sigma_1 \\ 
s(\epsilon) &= \epsilon\otimes \sigma_3, \quad  
s(a) = a\otimes \sigma_3 
. \tag2.8
\endalign
$$
In particular, the formulas
$d(v)=[\epsilon,v]_+$ and $d(a)=[\epsilon,a]_+$ mean
$$     \align
d(s(v))&=[s(\epsilon),s(v)]_+ = [\epsilon,v]\otimes \sigma_3\sigma_1 \\ 
d(s(a))&=[s(\epsilon),s(a)]_+ = [\epsilon,a]_+\otimes \sigma_3^2=[\epsilon,
a]_+\otimes 1.\endalign
$$
where the sigma matrices account for the correct signs. We stress
that the auxiliary space is just a tool to keep track of the
signs (the even and odd case could actually be handled in a
unified manner, but we choose to utilize $\Gamma$ in the even
case).

In both cases, {\it the symbols $x=v,\delta,\epsilon$ and $a$ in (2.6)
should actually be interpreted as $s(x)$ as specified above, and only
for simplicity of the notation we write $x$ instead of $s(x)$.} 
We will sometimes also use this simplified notation below, in particular 
in the Appendix, but we will always clearly point this out.

In the standard discussion of anomalies in quantum field theory one
constructs cocycles $c_{n,k}$ in the $(\Omega_*, \delta)$ complex by
integrating de Rham forms $\omega_{n,k}$ in $\Omega_*^n$ over a
compact manifold of dimension $n.$ In the NCG setting integration of
forms is replaced by applying an appropriate trace functional to the
operator valued forms. In fact, in the case of an odd Fredholm module
the integral is  normally  defined as the trace
$\tr_C \phi$ where the conditional trace is defined as
$\tr_C(X)=\frac12 \tr(X+\epsilon X\epsilon).$ However, in our notation
we have to take the trace also in the auxiliary space $\Bbb C^2,$ and the
correct definition is
$$ \int \phi= \frac12 \tr_C \,\sigma_3\phi.\tag2.9$$
In the case of an even
Fredholm module the standard definition of the integral is
$\tr_C(\Gamma\phi)$ 
whereas with our conventions the operator expression $\phi$ is already
equipped (for odd $k$) with the $\Gamma$ factor and we set
$$\int\phi= \tr_C \phi.\tag2.10$$
In the case of even $k$ (in an even module) the integral vanishes. 

The
translation from the classical to NCG setting is straight-forward.
The de Rham exterior derivation is replaced by the operation $d$ 
described above (to the forms after the symbol '$\int$').  All the formal
manipulations are done exactly in the same way as in the classical 
BRST complex. Of course, the nontriviality of the cohomology classes 
depends on what is meant by the trace (and the definition of the trace
is intertwined by the choice of $(H, D_0, \Cal B)$ ). 

The construction of a family of cocycles $c_{j,k}$ over a Fredholm
module starts from the (NCG) operator valued Chern class
$F^n,$ 
where $F=d(a)+a^2\in\Omega^2$ is the curvature.  
The Chern-Simons form is defined by 
$$c_{2n-1,0}(a) = \int n\int_0^1 dt a [ t d(a) + t^2 a^2 ] ^{n-1} ,$$
where $t d(a) + t^2 a^2 = F(t)$ is the curvature associated with $ta$
(a path connecting $a$ and $0$). The other terms in the complex
(starting from the Chern class) are given by similar formulas where 
one has the curvature associated with $d+\delta$ and a path of vector
potentials connecting $a+v$ and $0$, 
$$c_{2n-k-1,k} = \int n\int_0^1 dt( a [ td(a) + t^2 a^2 + (1-t)d(v) ]
^{n-1})|_{[k]},
\quad k=0,...,n-1 
$$ where now the curvature in the square brackets is obtained starting
from the vector potential $ta+v,$ and $(\cdots)|_{[k]}$ means the
projection onto those terms which are of degree $k$ in the 'ghost' $v$, 
and similarly,
$$
c_{2n-k-1,k} = \int n\int_0^1 dt( v  
[td(v)+(t^2-t)v^2 ] ^{n-1})|_{[k]},
\quad k=n,\ldots, 2n 
$$
where now the curvature is associated with $t v.$ We stress that in
the previous formulas for $c_{2n-k-1,k}$, we use the simplified notation
where $x=v,a$ is short for $s(x)$ as discussed above. (We also recall
that one can obtain different but equivalent formulas for
$c_{2n-k-1,k}$ depending on the choice of path $a+v\to 0$, and we find
it convenient to use the path consisting of two straight lines $a+v\to
v$ and $v\to 0$ leading to cocycles with lowest possible powers of $a$
[12].)

The crucial property are the cocycle relations $\delta c_{j,k}=0$ for
all $j$ and $k$. In the standard case these are a consequence of the
fact that the cocycles are integrals of traces of matrix valued de Rham forms,
$c_{j,k}=\int_M \tr\,\omega_{j,k}$, which are linked by the so-called descent
equations. These equations start with a relations connecting the Chern
class and the Chern-Simons form, $F^n =
d\omega_{2n-1,0}+(\cdots)$, and then continue,
$$
\delta(\omega_{2n-1-k,k}) + d(\omega_{2n-1,k+1}) = (\cdots), 
\quad k=0,\ldots, 2n-1   
$$
where $(\cdots)$ are terms which have zero trace; they are explicitly
commutators of matrix forms.  Since all
but the first term in the l.h.s.\ in these latter equations vanish
under the combined trace and integral 
(the second term on the l.h.s.\ vanishes due to
Stokes theorem) and $\delta\int =
\int\delta$, one obtains the cocyle relations. In fact, in the
standard case one usually applies to these descent equations the matrix
trace, and therefore the terms $(\cdots)$ on the r.h.s.\ become zero.
In the NCG generalizations, an analog of the separate matrix
trace is usually not available (the matrix trace and the ordinary
integration are combined in the abstract integral), 
and it is therefore natural to also
keep the terms $(\cdots)$ [6]. In fact, as we will show, the
terms $(\cdots)$ contain interesting information: in case of the
double complex 
based in a spectral triple they allow to connect the forms
$\omega_{j,k}$ and $\omega_{j+2,k}$ in different dimensions ($d$
degree) but identical form degree, which, for example, provides an
important link of the Schwinger terms in different dimensions (Theorem
A4) and, in general, will allow us to obtain local forms of the NCG
cocyles (Theorem 3.1).

As an example, the second Chern form is in the classical case the
matrix trace of $F^2,$ where $F = d(A) +A^2$ is the curvature form
associated to a connection $A=A^{\mu}dx_{\mu}.$ 
In general, one then can check by straightforward
algebraic manipulations 
$$
F^2 = d(\omega_{3,0}) + [a, \tilde\omega_{3,0}]_+ 
$$
where $\omega_{3,0} = \frac12[d(a),a]_+ +\frac23 a^3$ is the three
dimensional Chern-Simons form (before applying the trace functional)
and $\tilde{\omega}_{3,0} =
\frac13[d(a),a]_+ +\frac12 a^3$ is another 3-form. Note that for 
standard de Rham forms, the second term vanishes under the matrix
trace. If we apply this the relation in the Fredholm module setting we use $\int
F^2 = \tr_C(\Gamma F^2)$, and since $\Gamma [a, \tilde\omega]_+ =
[\Gamma a,\tilde\omega]$ the second term becomes a trace of a commutator
which vanishes for appropriate $p$. Of course, in the classical case also
the integral of the first term is zero in the case of a manifold without
boundary. The other descent equations
are, in the simplified notation where $x=a,v,\omega_{3-k,k}$ etc.\ is  
short for $s(x)$ as in Eqs.\ (2.7)-(2.8), 
$$
\align 
\delta(\omega_{3,0} ) + d(\omega_{2,1}) &= -\frac13[a,\omega_{2,1}]_+ 
-[v,\omega_{3,0}]_+  \\
\delta(\omega_{2,1} ) + d(\omega_{1,2}) &= -\frac23 [a,\omega_{1,2}]_+ 
-[v,\omega_{2,1}]_+  \\
\delta(\omega_{1,2} ) + d(\omega_{0,3}) &= -[v,\omega_{1,2}]_+ \\ 
\delta(\omega_{0,3} ) &= -\frac 12[v, \omega_{0,3}]_+\endalign  
$$
where
$$
\omega_{2,1} = \frac12[a,d(v)]_+, \quad \omega_{1,2} = \frac12 [ d(v),v]_+, 
\quad  \omega_{0,3} = -\frac 13 v^3 ,  
$$
all of them can be checked by straightforward algebraic manipulations.
As we have already said, the BRST ghost $v$ can be interpreted as the
Maurer-Cartan form on a group manifold, and thus are to be evaluated along
tangent vectors $X_j$ at the neutral element (i.e, Lie algebra
elements). Moreover, in the end we are interested in integrals of
the operator forms and we need the treat the cases of odd or even
Fredholm modules separately, as discussed above.

For example, the form $\omega_{1,2}$ when evaluated for Lie algebra elements
$X,Y$ and integrated in an odd module of appropriate Schatten index $p$, 
gives
$$
\int \omega_{1,2}(X,Y) = \frac12 \tr_C\, ([d(X),Y]_+ -  [d(Y),X]_+),$$
where $d(X)=[\epsilon,X],$ and $\omega_{0,3}$ leads to
$$ \int \omega_{0,3}(X,Y,Z) = -  \tr_C [[X,Y],Z]_+ , 
$$
and $\omega_{2,1}$ in an even module (appropriate $p$) leads to
$$ \int \omega_{2,1}(X)= -\frac12 \tr_C [a, \Gamma d(X)]_+
=\frac12\tr_C \Gamma [a,d(X)].$$

\vskip 0.3in
3. 'LOCAL'  NCG ANOMALIES AND SCHWINGER TERMS  

\vskip 0.3in
In the case of the classical BRST complex all the cocycles
$c_{j,k}$ are given in terms of differential forms which are 
differential polynomials in variables $a,v.$ The Fredholm module 
cocycles involve terms like $[\epsilon, v], \epsilon a+ a\epsilon,$ 
and therefore are nonlocal in nature; when evaluated using the symbol 
calculus of pseudodifferential operators they contain terms of
arbitrary high order in the partial derivatives. 

However, even in the case of the Fredholm module cocycles (for
classical vector potentials and gauge transformations) the locality is
preserved in a certain sense. Namely, it turns out that the cocycles 
$c_{j,k}$ are equivalent (in the BRST cohomology) to cocycles 
$c'_{j,k}$ which can be written as renormalized traces of
commutators of PSDO's. In the case of Schwinger terms this was
observed in [8] and the same principle was applied to the calculation
of the gauge anomaly for the chiral Dirac determinant in [9].  
The trace of a commutator depends only on the term in the asymptotic 
expansion of a PSDO which has order equal to $-\text{dim}M,$ and for this reason 
one needs to take into account only a finite number of derivatives 
of the symbols (since each differentiation in a homogeneous term 
decreases the order by one). In this sense the trace of a commutator
is a local expression.  In a more general setup, beyond the PSDO
calculus, we take this as a \it definition \rm of locality: cocycles 
which are traces of commutators in the algebra are called local. 

We set up the following assumptions.  There is a complex linear
functional TR on the algebra generated by $D_0,|D_0|$ and
$\Cal B$ such that 1) it is equal to the
ordinary trace for trace class operators, 2) it has the property that 
$\TR[A,B]=0$ when $AB,BA\in L_{1+}$ , and 3) $\TR\,
[\epsilon,W]=0$ (odd case), $\TR\, [\Gamma \epsilon,W]=0$ (even case),  
for bounded operators  $W,$ with $\epsilon=D_0/|D_0|.$ 

\bf Example 1 \rm Set $p=1$ and consider the cocycle $c(X,Y)= \tr_C\,
X[\epsilon,Y],$ for $X,Y\in\Cal B.$ Here everything  is defined in the 
original Hilbert space $H$ and not in $H\otimes \Bbb C^2.$ 
 We can write 
$$\align c_{1,2}(X,Y)&= \frac14  \tr\,\epsilon[\epsilon,X][\epsilon,Y]= \tr_C\,X[\epsilon,Y]\\
&=\frac12 \TR \, X[\epsilon,Y]= \frac12\TR\, [X\epsilon, Y] -
\frac12\TR\,\epsilon[X,Y].\endalign $$ 
The last term is the coboundary of the cochain
$\theta(X)=\frac12 \TR\,\epsilon X$ and therefore the class of $c_{1,2}$ is given by 
$$c_{loc} (X,Y) = \frac12\TR\, [X\epsilon, Y].$$ 
One can also check by a direct computation that $c_{loc}$ is a
cocycle. 
$$\align 2(\delta c_{loc})(X,Y,Z)&= \TR\,\{ [[X,Y]\epsilon,Z] + cycl.\}\\ 
&= \TR\, \{ [[X,Y], [\epsilon,Z]] + cycl. \} \\
&= \TR\, [\epsilon, [[X,Y],Z]] +\TR\, [Y,[[\epsilon,X],Z]] 
-\TR\, [X,[[\epsilon,Y],Z]]+cycl.\endalign$$
By 3) the first term on the right vanishes and by 2) the second and
third term vanish. In the case when $X,Y$ are multiplication
operators, by smooth functions, on the circle $S^1$ the local cocycle
becomes the central term in an affine Lie algebra,
$$c_{loc}(X,Y) =\frac{1}{2\pi i}\int_{S^1} \tr\, X dY,$$
where the trace under the integral sign is a finite-dimensional matrix
trace for the matrix valued functions $X,Y.$ 

The cocycle $c_{1,2}$ (or $c_{loc}$) arises in canonical quantization
in the following way. To each pair of basis vectors $e_i,e_j$ in $H$
there corresponds an operator $\hat e_{ij}=a^*(e_i)a(e_j)$ in the Fock
representation. We may label the basis vectors such that $e_i\in H_+$
for $i=0,1,2,\dots$ and $e_i\in H_-$ for $i=-1,-2,\dots,$ where
$H_{\pm}=\frac12(1\pm \epsilon)H.$ Then a
matrix $(\alpha_{ij})$ in this basis has the canonical quantization as 
the operator $\hat\alpha=\sum \alpha_{ij} \hat e_{ij}.$ For infinite matrices
this might diverge. Actually, that happens already when $\alpha$ is
the unit matrix. To circumvent this one introduces the normal ordering 
$\hat e_{ij} \mapsto \hat e_{ij} - \delta_{ij} \theta(-i)$ with
$\theta(x)=x$ for $x\geq 0$ and $\theta(x)=0$ for $x<0.$ With these
new operators $\hat e_{ij}$ the operator $\hat \alpha$ is defined in a
dense domain for any bounded operator $\alpha$ such that
$[\epsilon,\alpha]$ is Hilbert-Schmidt and the commutation relations
are given by 
$$[\hat\alpha,\hat\beta]=\widehat{[\alpha,\beta]} + c(\alpha,\beta),$$
[21]. 

\bf Example 2 \rm Here we consider the problem arising from
quantization of gauge currents in three space dimensions. Typically,
$[\epsilon,X]$ is not Hilbert-Schmidt but it belongs to the ideal
$L_{3+}\subset L_4.$ For this reason the expression for the 2-cocycle in the
previous example does not converge for 3-dimensional gauge
currents. Instead, one has to introduce a renormalization of the
2-cocycle,
$$ c_{3,2}(a;X,Y)= \frac18\tr_C \, a[[\epsilon,X],[\epsilon,Y]],$$ 
with $a=F_A-\epsilon.$
One can check by a direct calculation that this is a cocycle in the
sense that 
$$ c_{3,2}(a; [X,Y],Z) +\delta_X c_{3,2}(a;Y,Z) +\text{ cyclic
perm. of $X,Y,Z$ } =0.$$ 
Let next $\eta(a;X)= \frac18 \TR\, \epsilon a[\epsilon,X].$
By a direct calculation one can check that 
$c_{3,2} = \delta\eta + c_{loc}$ where now 
$$8 c_{loc}(a;X,Y)= \TR[Y,\epsilon a[\epsilon,X]]
-\TR[X,\epsilon a[\epsilon, Y]] +2\TR[X\epsilon,Y] -2\TR
[Y\epsilon,X]$$   
is explicitly a generalized trace of a sum of commutators. Using the
'bare' BRST notation (without the auxiliary space $\Bbb C^2$)
we can write $c_{loc}= \frac14 \TR [v\epsilon,v]-\frac18
\TR [v,\epsilon a [\epsilon,v]]. $

In this example the canonical quantization of gauge currents is
ill-defined even after normal ordering, precisely because
$[\epsilon,X]$ is not Hilbert-Schmidt. However, there is an operator
theoretic interpretation for second quantized $\hat X,\hat Y.$ These
are now generators for unitary transformations between Fock spaces
carrying inequivalent representations of the CAR algebra.
Geometrically, there is a bundle of Fock spaces parametrized by the
external field $a$ and the gauge transformations act as unitary maps 
between the fibers, [22].

\bf Example 3 \rm As a final example we discuss the gauge 
anomaly in two space-time dimensions.  
Here we are in the even case and we have
$\Gamma\epsilon=-\epsilon\Gamma$ and we consider gauge transformations
$X$ which commute with $\Gamma.$ Then 
$c_{2,1}(a;X)= \tr_C \,\Gamma a[\epsilon,X]$ is a cocycle,
$$\delta_X c_{2,1}(a;Y) -\delta_Y c_{2,1}(a;X) -c_{2,1}(a;[X,Y])=0,$$
using the fact that $\tr_C[\epsilon,\cdot]=0.$ In this case 
$$c_{2,1}(a;X) = \TR[\Gamma a\epsilon,X] + (\delta\eta)(a),$$ 
where $\eta(a)= \TR\, \Gamma a\epsilon.$ 
In BRST notation, for an even module, $c_{loc}=
\TR [a\epsilon,s(v)].$ We leave it as an exercise
for the reader to show that when inserting $\epsilon= D_0/|D_0|,$ 
$D_0= -i\sum_{k=1}^2 \gamma^k \frac{\partial}{\partial x_k},$ 
$D_A= D_0 + \sum \gamma^k A_k(x),$ and $a= D_A/|D_A|-\epsilon$ one obtains 
the standard formula for the nonabelian gauge anomaly in two
space-time dimensions,
$$c_{loc} =  \frac{1}{4\pi} \int \tr\, (A_1\partial_2 X - A_2\partial_1 X),$$ 
for a smooth infinitesimal gauge transformation $X$ of compact support.

In the general case we have the following result:

\proclaim{Theorem 3.1} Let $n=1,2,\dots $ and $k=0,1,2,\dots$ and let
the cocycle $c_{2n-1-k,k}$ be computed from the descent equations in
Section 2. Then for even $k$ the cohomology class $[c_{2n-1-k,k}]$ 
is represented by a cocycle $c^{loc}_{2n-1-k,k}$ which is a generalized 
trace of a sum of commutators;  each commutator is 
a polynomial in the operators $a$, $\epsilon$, and $X_j$ (the latter
operators correspond to the tangent vectors at which the ghosts
where evaluated). 
In the case of odd $k$ one has to add a term proportional to a 
generalized trace of $\omega_{0,k}.$
\endproclaim

This theorem is a reformulation of Theorems A4 and A5  in the
Appendix, which also gives more explicit formulas.

\vskip 0.3in

4. A MODEL FOR $\Cal A/\Cal G$ 

\vskip 0.3in
When $\Cal A$ is the space of classical smooth vector potentials 
on a compact manifold $M$ and $\Cal G_0$ is the group of based smooth 
gauge transformations (based means that $g(p)=1$ at some given point 
$p\in M$) then the quotient $X=\Cal A/\Cal G_0$ is a smooth infinite 
dimensional Banach manifold. If one tries to generalize this to the
NCG setting, by replacing $\Cal A$ by all bounded perturbations of the 
free Dirac operator and taking $\Cal G$ as the group of unitaries in
the algebra $\Cal B,$ one encounters the problem that there is no
natural way to define what is meant by based gauge transformations;
this leads to the difficulty that $X$ is not a manifold, it has a lot of 
singularities since at a generic point in $\Cal A$ the action of $\Cal
G$ is not free.  Here we shall construct a model for $\Cal A$ and
$\Cal G$ such that the quotient  will be free of singularities. 

Our construction is essentially based on Bott periodicity. 
Recall that the inductive limit $U(\infty)$ of the group $SU(N)$ as
$N\mapsto\infty$ has odd homotopy type: All its homotopy groups
$\pi_{2k+1}U(\infty)$ are isomorphic to $\Bbb Z$ whereas the even
homotopy groups are trivial. For this reason the group of based gauge
transformations $f:M\to U(N),$ in the limit $N\mapsto\infty,$ for 
$M=S^{2n}$ is homotopic to $U(\infty).$ In the odd dimensional case,
$M=S^{2n+1},$  the group of gauge transformations has even homotopy
type: All the even homotopy groups are isomorphic to $\Bbb Z$ and the 
odd homotopy groups are trivial. Denoting the gauge group in the even case 
by $G$ (which has the homotopy type of $U(\infty)$) then the group of 
gauge transformations in the odd case is the group of based loops 
$\Omega G.$ This latter group has the homotopy type of $U'_p$ for any 
$p\geq 1$ where $U'_p$ is defined in the following way:  Let $\epsilon$ 
be a grading operator (it could be the sign of a Dirac operator) in a 
Hilbert space, with eigenvalues $\pm 1,$ both eigenspaces infinite
dimensional. Set $U'_p(H)=\{g\in U(H) | [\epsilon,g] \in L_p\}.$ Here
$U(H)$ is the (contractible) group of all unitaries in the Hilbert
space $H.$ Recall that $U_{p+}(H)$ is the group of unitary elements in
the algebra $\Cal B.$ All elements $g$ in $U_{p+}(H)$ satisfy
$[\epsilon,g]\in L_{p+}.$

Let now $D_p,$ $p$ an even integer, be a hermitean operator in $H$ such that $1/|D_p| \in
L_{p+}$ and let $\Gamma$ be a hermitean operator in $H$ such that 
$\Gamma^2=1$ and $\Gamma D_p = -D_p\Gamma.$ Let $D_{p+1} = i\Gamma \frac{d}{dt} +
D_p.$ This operator is self-adjoint in an appropriate dense domain in
the Hilbert space $\Cal H= L^2(S^1, H)$ and has the property $1/|D_{p+1}|
\in L_{(p+1)+}.$  

A generalized vector potential is defined as a hermitean time dependent bounded 
operator $A(t)$ in $H$. The 'Dirac operator' coupled to $A(t)$ is then 
$D_{p+1} + A(t).$ The vector potential can be split as $A=A_0 +A_1,$ where 
$A_0$ commutes with $\Gamma$  (the 'time component' of $A$) and $A_1$ 
anticommutes with $\Gamma$ (this is the 'space component' of $A$). The 
time dependent gauge transformations are smooth functions $g(t)$ with
values in the group $U_{p+}(H,\Gamma)$ of unitary operators $g\in
U_{p+}(H)$ such that 
$[\Gamma, g]=0.$ We can split $H=H_1\oplus
H_2$ to eigenspaces of $\Gamma$ corresponding to eigenvalues $\pm 1.$ 
Since $g\in U_{p+}(H,\Gamma)$ commutes with $\Gamma$ we can write $g$ as a direct sum 
of linear operators $g_i : H_i \to H_i,$ $i=1,2.$    
 
The group $U_{p+}(H,\Gamma)$ is a subgroup of the group $U'_{p+}(H,\Gamma)$ which consists of
unitary operators $g$ such that $[\epsilon,g] \in L_{p+}$ and $\Gamma
g=g\Gamma.$ These conditions mean that $g=g_1\oplus g_2$ with $g_2
-\epsilon g_1 \epsilon \in L_{p+}.$ Thus the group $U'_{p+}$ is parametrized by pairs of
unitary operators $(g_2,h)$ with $g_2$ an arbitrary unitary
operator in $H_2$ and  $h= g_2^{-1} \epsilon g_1 \epsilon$ an
unitary operator in $H_2$ such that $h -1 \in L_{p+};$ we denote the
group of these elements by $U^{p+}(H_2).$ Thus by Kuiper's
theorem $U'_{p+}(H,\Gamma)$ is homotopy equivalent to $U^{p+}(H_2).$ 
A similar result holds for the subgroup $U_{p+}\subset U'_{p+}.$ In this
case one chooses as parameters $g_2$ and $\tilde h= g_2^{-1}
D_+^{-1} g_1 D_+$ and the conditions are as before. Here $D_+ : H_2\to 
H_1$ is the restriction of $D_p$ to $H_2.$ (If $D_+$ has zero modes,
replace $D_+^{-1}$ by $D_-/(D_-D_+ +1).$)

For each time dependent perturbation there is a unique (nonperiodic)
gauge transformation $g(t)$ such that $g(0)=1$ and $A'(t)= g^{-1} A g
+ g^{-1}[D_p,g]+ig^{-1}\partial_t g$ is the generalized temporal gauge,
i.e., the
even component $A'_0 =0.$  It follows that the quotient $\Cal A/\Cal
G_0,$ where $\Cal G_0$ is the group of periodic gauge transformations
$g(t)$ with $g(0)=1$, is equal to the product $\Cal A_1\times
U_{p+}(H,\Gamma).$ Here $\Cal A_1$ is the space of bounded operators $A$
in $H$ such that $\Gamma A= -\Gamma A$ and the coordinate $g\in
U_{p+}(H,\Gamma)$ comes from the holonomy $g=g(2\pi)$ around the circle. 

Since $\Cal A_1$ is an affine space, $\Cal A/\Cal G_0$ is homotopy 
equivalent to $U_{p+}(H,\Gamma).$  

We end this section by a remark on the homotopy type of the various
groups involved. 
When the condition $[\epsilon,g]\in L_{p+}$ is replaced by
$[\epsilon,g]\in L_p$ we obtain the group $U'_p(H,\Gamma) \subset
U'_{p+}(H,\Gamma).$ Note that also $U_{p+}\subset U'_{p+\alpha}$ for any 
$\alpha >0.$ Similarly one can define $U^p$ with $U^p\subset U^{p+}
\subset U^{p+\alpha}.$ According to Palais [13], the groups $U^p$
for all $p\geq 1$ are homotopy equivalent with $U(\infty).$ Similarly 
$U'_p(H)$ is homotopy equivalent to the group  $U'_0(H)$ of unitary
operators $g$ such that $[\epsilon,g]$ is of finite rank. 
It is plausible that the same holds for the groups $U_{p+}$ and
$U^{p+},$ but the proof in [13] cannot directly be applied to
these cases. The topology of these groups is determined by a norm 
topology on the operator ideals $L_{p+}.$ 

The natural norm on $L_{p+}$ 
for $p>1$ is given as
$$||T||_{p+} = \sup_{N\geq 1} N^{\frac1p -1} \sigma_N(T)$$
where $\sigma_N(T)$ is the sum of the $N$ largest eigenvalues of
$|T|.$ In the case $p=1$ the factor $N^{\frac1p -1}$ is replaced by 
$(\log N)^{-1}.$   
 
\vskip 0.3in
5. THE GERBE OVER $\Cal A/\Cal G_0$ 

\vskip 0.3in
Each $A\in\Cal A$ and $\lambda\notin Spec(D_A)$ defines a fermionic
Fock space $\Cal F_{A,\lambda}$ with a Dirac vacuum $|A,\lambda>.$ 
To begin with, we have the polarization $\Cal H= \Cal
H_+(A,\lambda)\oplus \Cal H_-(A,\lambda)$ to a pair of infinite dimensional 
subspaces, defined by the spectral projections $D_A > \lambda$ and 
$D_A < \lambda.$  Then the Fock space $\Cal F_{A,\lambda}$ is
generated by the algebra of creation and annihilation operators
$a^*(u),a(u)$ with the relations (2.1) 
and the characterization of the
vacuum $|A,\lambda>$ as in (2.2).

The Fock spaces depend on the choice of the vacuum level $\lambda.$
However, for $\lambda,\mu \notin Spec(D_A)$ there is a natural
projective isomorphism $\Cal F_{A,\lambda} \equiv \Cal F_{A,\mu}.$ 
This construction is equivariant with respect to the gauge group
action, leading to a projective bundle $P\Cal F$ over $\Cal A/\Cal
G_0.$ The question is whether there exists a true vector bundle 
$\Cal F$ over $X=\Cal A/\Cal G_0$ such that $P\Cal F$ is the
projectivization of $\Cal F.$   

In general, there is an obstruction to the existence of $\Cal F.$ The 
obstruction can be described in terms of an element $\omega$ in
$H^3(\Cal A/\Cal G_0, \Bbb Z).$  This may or may not correspond to 
a nontrivial de Rham cohomology class. However, in the present setting 
there is a nontrivial obstruction as a de Rham form.  

A more geometric way to describe the obstruction problem is to
construct a family of local complex line bundles
$DET_{\lambda\lambda'}$ over $\Cal U_{\l\l'}=\Cal U_{\l}\cap \Cal
U_{\l'}$ with $\Cal U_{\l}=\{A\in \Cal A/\Cal G_0 | \lambda\notin
Spec(D_A)\}.$ Here $DET_{\l\l'}$ is the top exterior power of the 
(finite dimensional) spectral subspace of $D_A$ corresponding to the 
open interval $(\l, \l')$ (with $\l <\l'$). These line bundles have a
set of natural isomorphisms 
$$DET_{\l\l'}\otimes DET_{\l'\l''} = DET_{\l\l''}$$ 
which give the relations needed to define \it a gerbe \rm over $\Cal
A/\Cal G_0.$ The gerbe is trivial if there is a family of local line 
bundles $DET_{\l}$ over the open sets $\Cal U_{\l}$ such that 
$DET_{\l\l'} = DET_{\l}^{-1} \otimes DET_{\l'}.$ Physically, these
latter bundles are the local fermionic vacuum bundles.  The
nontriviality of the gerbe is measured by the nontriviality of the 
\it Dixmier-Douady class \rm $\omega,$ [14].

The nontriviality of the obstruction as a de Rham form follows from
the considerations in [10, 15].
The setting in the latter paper was similar to the present case except 
that a smaller base space, $G_1$ of unitaries which differ from the identity by
a trace class operator, was considered instead of
$U_{p+}(H,\Gamma).$

On the group $G_1$ the obstruction form is particularly simple. It is given as
the left invariant form 
$$\omega(X,Y,Z)= \frac{i}{8\pi^2} \tr\, X[Y,Z]$$
where $X,Y,Z\in Lie(G_1).$ This is normalized such that its
integral over a fundamental 3-cycle in $U^3(H)\simeq U'_3(H,\Gamma)$
is equal to one.
As it stands, $\omega$  does not extend to $U'_p(H,\Gamma)$ for $p>3.$ 
Instead, we have to construct another representative for the cohomology class
which extends to $U'_p(H,\Gamma),$ and actually also to the larger group
$U_{p+}(H,\Gamma).$ Since $U_{q+}(H,\Gamma)\subset U'_p(H,\Gamma)$ for
$q <p,$ we get a normalized representative for the cohomology class also
in the former group.

We shall treat both the even and odd cases at the same time and $U'_p$ stands
for $U'_p(H,\Gamma)$ in the case of an even Fredholm module and $U_p=U'_p(H)$
in the odd case. In both cases we denote the left invariant Maurer-Cartan
1-form on the group by $g^{-1}\delta g.$  

Next we consider the $(d,\delta)$ BRST bicomplex on the group manifold
$U'_p.$ As before, $\delta$ is the exterior differentiation on the
group manifold with a choice of signs when acting on $d$ forms such
that $d\delta+\delta d=0.$ We set $\Delta=d+\delta,$ so $g^{-1}\Delta
g= g^{-1}[\epsilon,g] +g^{-1}\delta g$ and the second component is the 
Maurer-Cartan 1-form on the group, i.e., the BRST ghost $v.$

The form 
$$\theta_{k-j,j} = \int (g^{-1}\Delta g)^{k}|_{[j]}$$
is closed in the $\delta$ cohomology complex for any $k=0,1,2,\dots.$
Here $(\cdots)|_{[j]}$ denotes the component of $\delta$ degree 
(= ghost degree) $j$.
In order
that the integral is defined as a (graded) trace 
we have to assume that $k-j\geq p.$ 

Actually $\theta_{k-j,j}=0$ for all even $k$'s.
This  follows from the anticommutator relations 
for $d,\delta$ and from the cyclic properties  of the integral.  The closedness for odd $k$ follows from
$$\delta\theta= \int \delta (g^{-1}\Delta g)^k = \int \Delta (g^{-1} \Delta g)^k =
-\int (g^{-1}\Delta g)^{k+1} =0,$$
since $k+1$ is even.  

For $j=k$ and with a proper normalization $\theta_{k,j}$ is the
generator of $H^k(U'_p,\Bbb Z)$ whereas for $j=0$ the integral gives the
Fredholm index of $P_+ g P_+,$ where $P_+=\frac12(1+\epsilon).$ 

The case $j=3$ is of interest to us.  In this case $k-j=2n$ is even, so $\theta_{2n,3}$ is an integral of 
an even $d$-form.  We check that this is a nontrivial cohomology class.  For that purpose, choose 
$H$ as the Hilbert space of square integrable sections in a tensor product of a Dirac spinor bundle
over the torus $T^{2n}=(S^1)^{2n}$ and a trivial $\Bbb C^N$
bundle with the natural $U(N)$ action in the fibers.
The gauge transformations $g:T^{2n}\to U(N)$ act as multiplication operators on the spinor fields.
Choosing $D_0$ as the Dirac operator defined by the metric on the torus and the trivial vector potential 
$A=0$ in the $\Bbb C^N$ bundle, one has, [5,16],
$$\tr_C \, \Gamma a_0[\epsilon,a_1]\dots [\epsilon, a_{2n}]
=\frac{1}{n! (2\pi i)^{n}} \int_{T^{2n}} \tr\, a_0 da_1 \dots da_{2n},$$ 
for smooth functions $a_j:T^{2n} \to\gm,$ where $\gm$ is the Lie algebra of $U(N)$ in its fundamental 
representation and $'d'$ on the right means the de Rham exterior
derivative. 
It follows that, for any map $g(\cdot): S^3\to Map(T^{2n}, U(N)),$ we have
$$\int_{S^3} \theta_{2n,3} = \frac{1}{n!(2\pi i)^{n}}  \int_{S^3\times T^{2n}} \tr\, (g^{-1}dg)^{2n+3},$$ 
where $g$ is thought of as a $U(N)$ valued function on $S^3\times T^{2n}.$
In particular, for large $N,$ this integral is equal to
$24\pi^2\times$ a nonzero integer when $g$ represents a 
nontrivial homotopy class in $\pi_{3+2n}(U(N)) \simeq \Bbb Z.$ 
Since $\theta_{2n,3}$ is nontrivial on the subgroup of smooth gauge transformations in $U_{2n+}(H,\Gamma)$ 
it must represent a nonzero cohomology class of the latter group and therefore also in the moduli
space $X=\Cal A/\Cal G_0.$ 

Specializing from the result of Theorem 3.1 (cf. Lemma A3 in the
appendix) to the flat case
$a=g^{-1}[\epsilon, g]$ we observe that the forms $\theta_{2n,3}$ 
can be written as sums of generalized traces of
commutators plus a   properly 
normalized 3-form $\TR\, (g^{-1}\delta g)^3,$ modulo exact forms.  

\vskip 0.3in

6. SCHWINGER TERMS FROM THE DIXMIER-DOUADY CLASS 

\vskip 0.3in

In this section we shall work with the Fredholm modules based on the
ideals $L_p$ instead of $L_{p+},$ and likewise with the groups of the
type $U'_p.$ This is because we want to use available information
(based on results in [13]) on the homotopy type of the infinite
dimensional groups. 

Let $\theta_3$ be a closed integral 3-form on $X=\Cal A/\Cal G_0.$ The
pull back $\pi^*\theta_3=\psi_3$ with respect to the canonical
projection is a closed 3-form on the contractible space $\Cal A$ and
therefore $\psi_3= d\psi_2$ for some 2-form $\psi_2.$ (In this section
$d$ means always the de Rham exterior derivative.) 

Let $\Cal U_{\alpha}$ be an open contractible subset of $X$ and write
$\theta_3= d\theta_{2, \alpha}$ on $\Cal U_{\alpha}.$ We define 
$$\eta_{\alpha}= \psi_2 - \pi^*\theta_{2,\alpha}$$ 
on $\pi^{-1}(\Cal U_{\alpha}).$ 
Now $d\eta_{\alpha} = \psi_3 -d\pi^*\theta_{2,\alpha} = 
\psi_3 - \pi^*d\theta_{2,\alpha}=\psi_3 -\pi^*\theta_3 =0.$

If $\Cal U_{\beta}$ is another open subset of $X$ then on $\Cal U_{\alpha}\cap
\Cal U_{\beta}$ we have $\eta_{\alpha} -\eta_{\beta} =
\pi^*(\theta_{2,\alpha} - \theta_{2,\beta}).$  From this follows, by
the definition of the pull back map, that $\eta_{\alpha}$ and
$\eta_{\beta}$ agree \it when evaluated along gauge orbits. \rm  
Thus there is a  vertical 2-form $\eta$  on $\Cal A,$ defined everywhere 
on $\Cal A,$ but which is closed only along gauge directions.   

The vertical form $\eta$ defines an extension of the Lie algebra of 
$\Cal G$ by the abelian Lie algebra $\bold a$  consisting of functions 
$f:\Cal A\to\Bbb C.$  The commutators are defined by
$$[(X,f),(Y,g)] = ([X,Y], \delta_X g -\delta_Y f +\eta(X,Y)).$$ 
On the right the Lie algebra elements $X,Y$ are interpreted as
vertical vector fields on $\Cal A.$ The Jacobi identity is precisely 
the condition that $\eta$ is closed along vertical directions.

We can compute $\eta$ more explicitly. 
Let $\Cal P G$ denote the group of smooth paths in a group $G$ originating
from the identity in $G.$ The product is defined as a pointwise
multiplication along paths.
 Any $A\in\Cal A$ splits
uniquely as $A=A_0+A_1$ to even and odd components with respect to
$\Gamma.$ The even component $A_0$ is equal to $ig(t)^{-1} \partial_t
g(t)$ for a uniquely defined smooth function $g:[0,1]\to U'_{p}(\Gamma,
H)$ with $g(0)=1.$ As in Section 4, we can write $\Cal A= \Cal A_1 \times
\Cal P U'_{p}(\Gamma, H)$ and $\Cal A/\Cal G_0 = \Cal A_1 \times
U'_{p}(\Gamma, H).$  

Since $\Cal A_1$ is a vector space, any closed form on $\Cal A_1$ is
exact and we may assume without loss of generality that $\theta_3$ is
a pull-back of a form on $U'_{p}(\Gamma,H)$ with respect to the
projection on the second factor. 

By a direct calculation one can check that $d\psi_2=\pi^*\theta_3$
when $\psi_2$ is defined as 
$$\psi_2(u,v)= \int_0^1 \theta_3(g^{-1}\partial_t g, u_0(t),v_0(t))dt,$$ 
where the components $u_0,v_0$ of the tangent vectors $u,v\in \Cal A$ 
are given as paths in the Lie algebra of $U'_p(\Gamma,H).$ 
The along vertical directions, $\eta$ is equal to $\psi_2$ and thus 
$$\eta(g; X,Y)=  \int_0^1 \theta_3(g^{-1}\partial_t g, X(t), Y(t))dt,$$ 
for periodic functions $X(t),Y(t)$ with values in
$Lie(U'_p(\Gamma,H)).$  

Since $\pi_2(U'_p(\Gamma,H))=0,$ the group $\Cal G_0$ of (based) loops 
in $U'_p(\Gamma,H)$ is simply connected. On the other hand, $\pi_2(\Cal L_k
U'_p(\Gamma, H))=\pi_3(U'_p(\Gamma, H)) =\Bbb Z$ for any of the
connected components $\Cal L_k$ of the group $\Cal LU'_p$ of smooth loops in
$U'_p.$
By Hurewicz' theorem,
$H^2(\Cal L_k U'_p(\Gamma,H), \Bbb Z) = \Bbb Z.$  In fact, the 2-form $\eta$ 
restricted to a gauge orbit gives a representative for the basic class
in $H^2(\Cal L_kU'_p,\Bbb Z).$  

We shall work only in the connected component
of identity $\Cal L_0 U'_p.$  
The group extension of $\Cal L_0 U'_p(\Gamma,H)$ corresponding to the Lie
algebra extension above can be explicitly constructed in the same way
as was done in [17] in the case of groups of classical gauge
transformations. 

The construction starts from the Cartesian product $\Cal P (\Cal L_0U'_p)\times
Map(\Cal A, S^1).$ The product
is defined as 
$$(f_1, \alpha_1)(f_2,\alpha_2) = (f_1f_2, \alpha_1 \alpha_2^{f_1} 
\exp(2\pi i\int_K \eta),$$ 
where $f_1f_2$ is just the point-wise product in the path group,
$\alpha^f$ is the function obtained form $\alpha$ by gauge
transforming (using the right action) the argument by $f(1)\in \Cal L_0 U'_p.$ 
The phase is evaluated by integrating the vertical 2-form $\eta$ over 
a 2-surface $K$ in the fiber at $A\in\Cal A.$ The surface is chosen
such that its boundary consists of the paths $A\cdot f_1(t),$ $A\cdot
f_1(1)f_2(t),$ and $A\cdot f_1(t)f_2(t).$ The value of the phase
factor does not depend on the choice of $K$ since the difference in
the exponent is $2\pi i$ times an integral of the integral 3-form $\omega_3$ over a 
closed 3-surface in $U'_p$ and this is an integer. 
 
The extension $\widehat{\Cal L_0U'_p}$ is now defined as 
$$ \Cal P(\Cal L_0U'_p)\times Map(\Cal A, S^1)/\Cal Q,$$ 
where $\Cal Q$ is the group of based loops (at the identity) in $\Cal L_0U'_p$
and the right action of $\Cal Q$ is defined as the point-wise right
action on $\Cal P(\Cal L_0U'_p)$ and as the simultaneous phase shift
$\alpha\to\alpha',$ 
$$\alpha'(A)=\alpha(A) \exp(2\pi i \int_{\Sigma} \eta ),$$ 
where $\Sigma$ is the  2-surface in the fiber through $A$, bounded by
the loop  $A\cdot g(t),$ $g\in\Cal Q.$  

\vskip 0.3in

7. THE CASE OF A NONCOMMUTATIVE TORUS 

\vskip 0.3in

In this section we want to apply the general results for computing
cocycles on a noncommutative torus. We start by recalling the basic
definitions, [18].  

Consider an algebra consisting of finite linear combinations of
'noncommutative Fourier modes'  $u^r,$ where $r=(r_1,\dots,r_n)$ is a
vector of integers, 
$$u^r= e^{\pi i \sum_{j<k} \theta_{jk}r_j r_k} u_1^{r_1}\dots
u_n^{r_n}$$ 
and the relations
in the algebra are given as 
$$u_i u_j = e^{-2\pi i \theta_{ij} } u_j u_i, \tag7.1$$ 
where $\theta$ is a real antisymmetric matrix.  These relations lead
to the multiplication formula 
$$u^r \cdot u^{r'}= \lambda(r,r') u^{r+r'} \tag7.2$$ 
with $\lambda(r,r')=\exp(-\pi i\sum  r_j\theta_{jk} r'_k).$ 

There is an antilinear conjugation defined by $(u^r)^* = u^{-r}$ and a 
trace $\tau$ such that $\tau(u^r)=0$ if $r\neq 0$ and
$\tau(u^0)=\tau(1) =1.$  We define a Hilbert space $H$ by tensoring
the above algebra with the spinor and internal symmetry space 
$V=\Bbb C^{[(n+1)/2]}\otimes \Bbb C^N$ and completing with
respect to the inner product defined by $<u^r,u^{r'}> = \tau((u^r)^*
u^{r'}) = \delta_{rr'}.$ 
Here $[a]$ is the integer part of a real
number $a.$ 
The trace is extended to the tensor product
of the torus algebra and of End$V$ by the matrix trace on End$V.$ 
The adjoint of a multiplication operator by the element
$u^r$ is  $(u^r)^*= u^{-r}.$  

There is a commutative family of derivations $\delta_j$ ($j=1,2,\dots,n$)
defined by $\delta_j (u^r)= r_j u^r.$  These derivations are
represented by unbounded operators $P_j$ in the Hilbert space $H$
through $P_j u^r = r_j u^r$ which means that $[P_j, u^r] = r_j u^r$ 
for the corresponding multiplication operators $u^r$ (we use the same
notation for elements in the noncommutative torus algebra and vectors
in the Hilbert space).   

The (free) Dirac operator is defined by 
$$D_0= \sum \gamma^k \delta_k,$$ 
where the gamma matrix algebra is defined using the standard Euclidean
metric, $\gamma_i\gamma_j+\gamma_j\gamma_i= 2\delta_{i,j}.$ 

As in the case of pseudodifferential operators on an ordinary torus we
can introduce a symbol calculus.  Consider first classical PSDO
symbols $f=f(p)$ which are functions of the momenta $p_j$ only and
which have an asymptotic expansion $f \sim f_N + f_{N-1} + \dots $ as a 
series of homogenenous symbols $f_j$ of order  $j$ in the momenta,
i.e, $f_j(\lambda p) = \lambda^j f(p)$ for $\lambda >1$ and for large 
momenta.  Each such a symbol defines an (in general unbounded)
operator in $H$ through $ f u^r = f(r) u^r.$  More generally, we
consider operators 
$$f = \sum_r   u^r f^{(r)} ,$$ 
where each $f^{(r)}$ is a classical symbol of the above type tensored 
with a matrix operating in $V.$ The definition makes
sense even for infinite linear combinations assuming that the sequence
$|f^{(r)}(p)|$ is rapidly decreasing as $|r|\to\infty$ for all values
of $p.$ From the definitions
follows that 
$$ u^{-r} f(p) u^r = f(p+r)\tag7.3$$ 
and therefore the product of a pair $f,g$ of operators is defined by
the 'star product' 
$$(f*g)^{(r)}(p)  = \sum_s  f^{(r-s)}(p) g^{(s)}(p+s)
\lambda(r-s,s).\tag7.4$$     
 
The only difference as compared to the commutative torus is the
appearance of complex phases $\lambda$ on the right-hand side.  
The trace of the operator $f$ (when it is defined) is given by 
$$\tr\, f=  \sum_{p\in \Bbb Z^n} \tr\, f^{(0)}(p),$$ 
where the trace on the right is the ordinary matrix trace in $V.$
As usual, $f$ is trace-class if and only if ord$(f) < -n.$  
The border line case ord$(f) = -n$ is of special interest to us. 

For complex numbers $z$ with an enough large real part one can define 
$$\zeta(z;f) = \tr\, f (|P| +1)^{-z}.\tag7.5$$ 
If $f$ is trace-class then $\zeta$ is holomorphic at $z=0$ and $\tr\,
f =\zeta(0;f).$  If ord$(f)\geq  -n,$ the function $\zeta$ has in general 
a pole at $z=0.$ The pole depends only on the component $f_{-n}.$  
The proof of this statement follows from the corresponding result for 
PSDO's on the commutative torus since the spectrum of $f^{(0)}$ is
exactly the same as in the commutative case.   For the same reason 
the the residue at $z=0$ can be calculated as a momentum space
integral,

$$\text{Res}(f) = \text{res}_{z=0} \zeta(z;f)= \int_{|p|=1} \tr\,
f_{-n}^{(0)}(p) dp.\tag7.6$$

For classical PSDO's on a compact manifold there is a useful formula 
relating the renormalized (noncyclic) trace of a commutator to an
operator residue. This formula was used for calculation of Schwinger
terms in [8,19] and later generalized in [20] 
to a wider class of operators. First one defines a renormalized trace 
for nontraceclass PSDO's  by 
$$\TR\, A  = \lim_{z\to 0} (\tr\, A |D|^{-z} - \frac{1}{z} \Res{A}).\tag7.7$$ 
The lack of cyclicity of $\TR$ is given by 
$$\TR[A,B] = \Res\, A[\ell,B] ,\tag7.8$$ 
where $\ell= \log|D_0|=\log|P|.$ We check that the same formula holds in the
case of noncommutative torus.  By linearity, it is sufficient to prove
the formula for each Fourier mode separately. So we take $A=u^r f(p) $ 
and $B= u^s g(p) .$ By the definition of the trace, $\TR\, AB$ is
nonzero only when $s=-r,$ so we assume this. But now $\TR\, AB = \TR\,
\lambda(r,s) \tilde{f} g,$ where $\tilde f(p)= f(p-r).$ Since
$\lambda(r,s)=1$ when $s=-r,$ the trace formula for NC
torus reduces to the formula on commutative torus and thus (7.8)
holds.   

Let us look closer at the case $n=3.$  We have seen that the Schwinger
term in three dimensions is equivalent (modulo coboundaries) to the
'local' expression 
$$8\omega_{3,2} = \TR\,\{ [X, \epsilon a [\epsilon,Y]]-
[Y,\epsilon a [\epsilon,X]] -2[X\epsilon,Y] +2[Y\epsilon, X]\}.$$   
We now apply the residue formula in the case of noncommutative
multiplication operators $X,Y$ and we obtain 
$$8\omega_{3,2}= - \Res\, \epsilon a [\epsilon,Y] [\ell, X]
+\Res\, \epsilon a [\epsilon,X] [\ell, Y].\tag7.9$$

Taking $a=F-\epsilon$ and $F= (D_0+A)|D_0+A|^{-1},$ the sign of a Dirac
operator defined by a potential $A=\sigma^k A_k$ on the noncommutative 
torus, a straight-forward computation gives 
$$\omega_{3,2}=
\frac{i\pi}{6}\tau ( A [[D_0,X],[D_0,Y]] )
=\frac{i\pi}{3}\epsilon^{ijk}\tau\left(A_i
(\delta_j(X)\delta_k(Y)-\delta_j(Y)\delta_k(X))\right) .\tag7.10$$

We have used the expansion 
$$F= \epsilon + A/|P| - \epsilon A^k P_k /|P|^2 + O(1/|P|^2)$$
and the formulas 
$$\align [\epsilon, u^{(r)}] &= u^{(r)} ( u^{(-r)} \epsilon u^{(r)} -\epsilon) 
= u^{(r)} \left( \frac{(P_k +r_k)\sigma^k}{|P+r|} - \frac{P_k\sigma^k}{|P|}\right)\\ 
&= u^{(r)} \left(r_k\sigma^k/|P| -\epsilon P_k r^k/|P|^2 + O(1/|P|^2) \right)\endalign$$ 
and similarly 
$$[\ell, u^{(r)}] = u^{(r)} P_k r^k/|P|^2 +O(1/|P|^2).$$ 
The derivation of (7.10) is then completed by taking account
$$\tr\,\sigma_i\sigma_j\sigma_k =2i \epsilon_{ijk} \text{ and }
\Res \frac{p_ip_j}{|p|^5} = \frac{4\pi}{3}\delta_{i,j}.$$

\vskip 0.3in



\newpage
\noindent
APPENDIX: NONCOMMUTATIVE DESCENT EQUATIONS

\vskip 0.1in
\noindent
We assume linear operators $d$ and $\delta$ acting on polynomials
generated by $a$, $v$, $d(a)$, $d(v)$, $\delta(a)$, and $\delta(v)$,
such that $d^2=0=\delta^2=d\delta+\delta d.$ These operations are then 
uniquely fixed by the additional rules in (2.6) 
together with $\epsilon^2=\id$. As mentioned in the text, we will use
the simplified notation $x=a,\epsilon,v$ and $\delta$ short for $s(x)$
defined in (2.7) resp.\ (2.8) for even resp.\ odd Fredholm modules
throughout this Appendix except in the Remark at the end.

We now state and give a proof of the non-commutative descent equations.
We define
$$
\align
F(t) &= td(a)+t^2 a^2  + (1-t) d(v) \\
F'(t) &= (t^2-t)v^2 + t d(v) 
\endalign
$$
and 
$$
\omega_{2n-1} = \int_0^1dt\,\Psi_{2n-1}(t),\quad 
\tilde\omega_{2n-1} = \int_0^1dt\, t \Psi_{2n-1}(t),\quad 
$$
and similarly for $\omega_{2n-1}'$ and $\tilde\omega_{2n-1}'$, where
$$
\align
 \Psi_{2n-1}(t) &= \sum_{\nu=0}^{n-1} F(t)^{n-1-\nu}
a F(t)^{\nu}  \\
\Psi'_{2n-1}(t) &= \sum_{\nu=0}^{n-1} F'(t)^{n-1-\nu}
vF'(t)^{\nu} . 
\endalign
$$


The non-commutative descent equations can be summarized by the following
algebraic identities: 

\proclaim{Lemma A1}
$$
\align
 \delta(\omega_{2n-1})+ d(\omega_{2n-1}) &= -[v,\omega_{2n-1}]_+
-[a ,\tilde\omega_{2n-1}]_+ + F^n - (d(v))^n \\ 
\delta(\omega'_{2n-1}) + d(\omega'_{2n-1}) 
&= -[v,\tilde\omega'_{2n-1}]_+ + (d(v))^n \tag A.2
\endalign
$$ 
where $F=d(a)+a^2$. \endproclaim
Comparing equal powers of $v$ we obtain from (A.2) 
$$
\align
d(\omega_{2n-1,0}) &= -[a ,\tilde\omega_{2n-1,0}]_+ + F^n 
\\ 
\delta(\omega_{2n-k,k-1})+ d(\omega_{2n-1-k,k}) 
 &=
-[v,\omega_{2n-k,k-1}]_+-[a,\tilde\omega_{2n-k-1,k}]_+ 
\\ & \quad\quad
k=1,\ldots n-1 \\ 
\delta(\omega_{n,n-1})  + d(\omega_{n-1,n})
 &=
-[v,\omega_{n,n-1}]_+ \tag A.3
\\
\delta(\omega_{2n-k,k-1}) + d(\omega_{2n-k-1,k}) 
 &= -[v,\tilde\omega_{2n-k,k-1}]_+ \\ & \quad\quad 
k=n+1,\ldots 2n -1\\
\delta(\omega_{0,2n-1}) &= -[v,\tilde\omega_{0,2n-1}]_+
\endalign
$$ 
where the $k$ in $\omega_{2n-k-1,k}$ 
is the ghost degree i.e. power in $v$,
and
$$
\align
\omega_{2n-1} &= 
\omega_{2n-1,0} + \omega_{2n-2,1} + 
\ldots + \omega_{n,n-1} + \ldots \\
\omega'_{2n-1} &= 
\omega_{n-1,n} + \omega_{n-2,n+1} + 
\ldots + \omega_{0,2n-1} + \ldots  
\endalign
$$
(note that the equation for $k=n$ is obtained by combining
the two equations
$$
\align
\delta(\omega_{n,n-1})  
 &= 
-[v,\omega_{n,n-1}]_+ -(d(v))^n 
\\
d(\omega_{n-1,n}) &=  (d(v))^n 
\endalign
$$
obtained from the first and the second relation in Eq.\ (A.2), 
respectively).

\bigskip
\noindent \demo{ Proof of Eq.\ (A.2)}  
We observe that
$$
F(t) = (\epsilon + \delta + t a + v)^2-\id
$$
implying 
$
\partial_t (F(t) )  = [ \epsilon + \delta + ta + v, a]_+ , 
$
and therefore
$$
\align
F^n-(d(v))^n & = \int_0^1 dt\, \partial_t( F(t) )^n  = \\
& \int_0^1 dt\,\sum_{\nu=0}^{n-1} F(t) ^{n-1-\nu}  
[ \epsilon + \delta + ta + v, a]_+ F(t)^{\nu} =  \\
&\int_0^1 dt\,[ \epsilon + \delta + ta + v, \Psi_{2n-1} ]_+   
\endalign
$$
(we used that $\epsilon+\delta+ta +v$ commutes with $F(t)$), 
implying the first identity in Eq.\ (A.2). In a similar manner, 
$$
F'(t) = (\epsilon + \delta + tv)^2-\id  
$$
implies $\partial_t(F'(t) ) =  [\epsilon + \delta + tv, v]_+$ and thus
$$
\align
(d(v))^n &= \int_0^1 dt\, \partial_t( F'(t) )^n  = \\
& \int_0^1 dt\,\sum_{\nu=0}^{n-1} F'(t) ^{n-1-\nu}  
[ \epsilon + \delta + tv, v]_+ F'(t)^{\nu} =  \\
&\int_0^1 dt\,[ \epsilon + \delta + tv, \Psi'_{2n-1} ]_+   
\endalign
$$
yielding the second identity in Eq.\ (A.2). 
{\QED}\enddemo

\vskip 0.1in
\noindent {\bf Some applications.}
A simple special case of Eqs.\ (A.3) is the following result.  We
point out that here, $x=a$ and $\epsilon$ {\it need not} to be
interpreted as $s(x)$ but can be taken directly as operators in $H$.

\proclaim{Lemma A2} For flat $a$, i.e., 
$a=g^{-1}[\epsilon,g]=[g^{-1}\epsilon,g]$ for some invertible operator $g$, 
the following holds true
$$
a^{2n+1} = 4a^{2n-1} + 2[\epsilon,a^{2n-1}\epsilon] - [a,\epsilon a^{2n-1}] 
\tag A.4 
$$
for all positive integers $n$.
\endproclaim

\noindent \demo{Proof}  
For flat $a$ we have $F=d(a)+a^2=0$, and thus the first
equation in (A.3), implies 
$
[\epsilon,\omega_{2n-1,0}]_+ = -[ a,\tilde\omega_{2n-1,0} ]_+ $
where $\omega_{2n-1,0}$ is obtained from $\omega_{2n-1}$ 
above by substituting $F(t) = (t^2-t)a^2$, i.e., 
$\omega_{2n-1,0} = \int_0^1 dt\, n(t^2-t)^{n-1} a^{2n-1}
\equiv c_n a^{2n-1}$,
and similarly 
$\tilde\omega_{2n-1,0} = \int_0^1 dt\, 
t n(t^2-t)^{n-1} a^{2n-1}\equiv \tilde c_n a^{2n-1}$. 
Note that 
$$
2\tilde c_n - c_n = \int_0^1 dt\, n(2t-1)(t^2-t)^{n-1} = 0  
$$
implying 
$$
2[\epsilon,a^{2n-1}]_+ = -[ a ,a^{2n-1} ]_+ .$$
Multiplying this identity by $\epsilon$ we obtain
$$
\align
2\epsilon [\epsilon,a^{2n-1} ]_+ &= 4 a^{2n-1} + 2[\epsilon,a^{2n-1}\epsilon] 
=\\
 - \epsilon [ a,a^{2n-1} ]_+ &= - [ \epsilon , a ]_+ a^{2n-1} + 
[a,\epsilon a^{2n-1}] \; . 
\endalign
$$
Inserting $[ \epsilon , a ]_+ = d(a) = -a^2$ we obtain Eq.\ (A.4). 
{\QED}\enddemo

Note that the Lemma gives as a byproduct a simple  proof of the known index
formula for the Fredholm index ind$P_+gP_+,$ where $g\in U'_p(H_+\oplus
H_-)$ and $P_+$ is the projection to $H_+.$ Namely, since according to
[13] the connected components of $U'_p$ are labelled by the Fredholm
index of $P_+gP_+,$  the index is a homotopy invariant, and the
multiplication operator by the function $f(x)= e^{inx}$ on the unit 
circle $0\leq x\leq 2\pi$ has
index ind$P_+ fP_+ =n$ (by a simple exercise in Fourier analysis), it
is sufficient to check that 
$$n= \frac12 \tr\, f^{-1}[\epsilon,f]$$ 
and the general formula 
$$\text{ind} P_+ g P_+ = 2^{-p}\tr\, (g^{-1}[\epsilon,g])^{p}$$ 
follows for odd positive integers $p.$ 
  
\vskip 0.1in
\noindent {\bf Local cocyles.} The basic result implying the existence
of the local cocycles is the following

\proclaim{Lemma A3} The forms defined above obey  
$$
\align
\epsilon[v,\omega_{2n-k,k-1}]_+ + \epsilon[a,\tilde\omega_{2n-k-1,k}]_+ &\simeq 
c_{n,k} \, \omega_{2n-k+1,k} , \quad k=1,\ldots,n-1 \\
\epsilon[v,\omega_{n,n-1}]_+ &\simeq  c_{n,n} \, \omega_{n+1,n}
\\
\epsilon [v,\tilde\omega_{2n-k,k-1}]_+ 
&\simeq  c_{n,k} \, \omega_{2n-k+1,k},\quad 
k=n+1,\ldots,2n \tag A.5
\endalign
$$
where
$$
c_{n,k}=\frac{2n+1-k}{n+1}
$$
and `$\simeq$' means `equal up to commutator terms'.
\endproclaim

\noindent
\demo{Proof}  
To prove the first relation in Eq.\ (A.5) we observe 
$$
\align
\epsilon[v,\omega_{2n-k,k-1}]_+ + \epsilon[a,\tilde\omega_{2n-k-1,k}]_+ 
&= [\epsilon,v]_+ \,\omega_{2n-k,k-1} + [\epsilon,a]_+ \,\tilde\omega_{2n-k-1,k}
+ \\ 
[\epsilon \omega_{2n-k,k-1} , v ] + [\epsilon\tilde\omega_{2n-k-1,k},a]
&\simeq d(v)\, \omega_{2n-k,k-1}
+ d(a)\, \tilde\omega_{2n-k-1,k}  
\endalign
$$
and thus what we actually need to prove is 
$$
d(v)\omega_{2n-k,k-1} + d(a)\tilde\omega_{2n-k-1,k} 
\simeq  \frac{2n+1-k}{n+1}\, \omega_{2n-k+1,k} . 
\tag A.6
$$ 
For that it is convenient to define an ordering symbol $\{\}$ so that 
$$
\sum_{N=0}^\infty 
(a_1+a_2+\ldots a_k)^N \, := \, \sum_{m_1,m_2,\ldots,m_k=0}^\infty \{
a_1^{m_1} , a_2^{m_2 },\ldots, a_k^{m_k}\} 
$$ 
for all operators $a_j$ (we regard this as generating function
defining all possible 
$\{a_1^{m_1} , a_2^{m_2 },\ldots, a_k^{m_k}\}$). 
We observe that this definition implies 
$$
a_1 \{a_1^{m_1} , a_2^{m_2 },\ldots, a_k^{m_k}\} 
\simeq \frac{m_1+1}{m_1+m_2+\ldots + m_k +1} 
\{a_1^{m_1+1} , a_2^{m_2 },\ldots, a_k^{m_k}\} . 
$$
We can thus write  
$\omega_{2n-k-1,k} = \int_0^1dt\, \Omega_{2n-k-1,k}$
and $\tilde\omega_{2n-k-1,k} = 
\int_0^1dt\, t\Omega_{2n-k-1,k}$ where 
$$
\Omega_{2n-k-1,k} = \{ u^k, (p+q)^{n-k-1},a\} = 
\sum_{\ell=0}^{n-k-1}  \{ u^k, p^{n-k-\ell-1},q^{\ell},a\} 
$$
with $u\equiv (1-t)d(v)$, $p\equiv td(a)$ and 
$q\equiv t^2a^2$. 
To prove Eq.\ (A.6) we thus need to show that 
$$
\align
&\frac{2n+1-k}{n+1}\int_0^1dt\,  \{ u^k, p^{n-k-\ell},q^{\ell},a\} 
\simeq \\
\int_0^1dt\, 
&\Bigl( \frac{u}{1-t}\{ u^{k-1}, p^{n-k-\ell},q^{\ell},a\} 
+ p \{ u^k, p^{n-k-\ell-1},q^{\ell},a\} 
\Bigr), 
\endalign
$$
which follows from 
$$
\align
\int_0^1dt\, \Bigr( \frac{2n+1-k}{n+1} -  
\frac{k}{(n+1)(1-t)}
-\frac{n-k-\ell}{n+1}  \Bigl)t^{n-k-\ell} t^{2\ell} (1-t)^k &\\   
= \frac{1}{n+1}\int_0^1dt\,\partial_t 
(t^{n+1-k+\ell} (1-t)^k)=0 . &
\endalign
$$

The proof of the second and third relations in Eq.\ (A.5) is
similar but simpler: Recalling $u=(1-t)d(v)$ we get 
$$
\align
\epsilon[v,\omega_{n,n-1}]_+ 
\simeq d(v) \omega_{n,n-1} 
= \int_0^1 dt\, \frac{1}{1-t} u [u^{n-1},a]_+  
= &
\\ \int_0^1 dt\, \frac{1}{1-t}  \frac{n}{n+1} [u^{n},a]_+ =   
\int_0^1 dt\, [u^{n},a]_+ = c_{n,n} \, \omega_{n+1,n} &
\endalign
$$
since
$$
 \frac{n}{n+1}\int_0^1dt\, (1-t)^{n-1} = \int_0^1dt\, (1-t)^{n}, 
$$
and finally, with $r\equiv td(v)$ and $s \equiv (t^2-t)v^2$
$$
\align
\epsilon [v,\tilde\omega_{2n-k,k-1}]_+ 
\simeq d(v)\tilde\omega_{2n-k,k-1}  
=\int_0^1 dt\, r \{r^{2n-k},s^{k-1-n},v\}  
\simeq & \\
\int_0^1 dt\,  \frac{2n+1-k}{n+1} \{r^{2n+1-k},s^{k-1-n},v\}  
=  c_{n,k} \, \omega_{2n-k+1,k} . &
\endalign
$$
{\QED}\enddemo

\bigskip 
We obtain a generalization of Lemma A2:

\proclaim{Theorem A4} Let
$$
\omprime_{2n-1-k,k} = C_{n,k} \, \omega_{2n-1-k,k} 
\hbox{\quad where \quad}
C_{n,k}=(-1)^n \frac{(2n-1-k)!!}{2^n n!}.
$$
Then
$$
\omprime_{2n+1-k,k} \simeq \omprime_{2n-1-k,k} - 
\frac{C_{n,k}}{2 C_{n,k-1}} \delta (\epsilon \omprime_{2n-k,k-1}). 
\tag A.7
$$
\endproclaim

\noindent \demo{Proof} Using the Lemmas above, we observe that
$$
-\delta(\epsilon \omega_{2n-k,k-1})+ 2\omega_{2n-1-k,k} \simeq
-c_{n,k}\omega_{2n+1-k,k}\quad k=1,2,\ldots, 2n-1, \tag A.8
$$
where we have used that 
$\epsilon d(\omega)= \epsilon[\epsilon,\omega]_+\simeq 2\omega$.
By multiplying this with $(1/2 )C_{n,k}$ and using the 
recursion relation
$$
-\frac12 c_{n,k}C_{n,k} = C_{n+1,k},
$$
we get Eq.\ (A.7). 
{\QED}\enddemo 

\bigskip 

\proclaim{Theorem A5} For even $k$ the $k$-cocyle
$$
\omloc_{2n+1-k,k} = \omega_{2n+1-k,k} + \delta B_{2n-k,k-1}
$$
is local for some cochain $B_{2n-k,k-1}$, i.e., a sum of commutator
terms. In the case when $k$ is odd one has to add a term proportional
to $\omega_{0,k}$ to the
right-hand-side of the equation.   
\endproclaim 

\noindent \demo{Proof}
Eq.\ (A.8) implies 
$$
\align
\omega_{2n+1-k,k} &\simeq -\frac{2}{c_{n,k}}\omega_{2n-1-k,k} -
\frac{1}{c_{n,k}} \delta(\omega_{2n-k,k-1}) \simeq \\ 
&\sum_{\ell'=0}^\ell (-1)^{\ell'+1}
\frac{2^{\ell'}}{c_{n,k}c_{n-1,k}\cdots c_{n-\ell',k}}
\delta(\omega_{2n-2\ell'-k,k-1}) \\ &+ 
(-1)^{\ell+1}
\frac{2^{\ell}}{c_{n,k}c_{n-1,k}\cdots c_{n-\ell,k}}
\omega_{2n-1-2\ell-k,k} . 
\endalign
$$
In particular, 
$$
\omega_{2n+1-2k,2k} \simeq \sum_{\ell'=0}^{n-k} (-1)^{\ell'+1}
\frac12 \frac{(n+1)!(n-2k-\ell')!}{(n-\ell')!(n-2k)!} 
\delta(\omega_{2n-2\ell'-2k,2k-1}) 
$$
and

$$
\align
\omega_{2n-2k,2k+1} & \simeq \sum_{\ell'=0}^{n-k-1}(-1)^{\ell'+1}
\frac12 \frac{(n+1)!(n-2k-\ell'-1)!}{(n-\ell')!(n-2k-1)!} 
\delta(\omega_{2n-2\ell'-2k-1,2k}) \\ & + 
(-1)^{n-k} \frac12 \left(\matrix n+1 \\  k\endmatrix \right)
\omega_{0,2k+1}  
\endalign
$$

which also gives an explicit formula for $B_{2n-k,k-1}.$ 
{\QED}\enddemo
 
\bigskip

\noindent {\bf Remark:} Theorem A5 does not yet imply Theorem 3.1: we recall 
that we worked on an abstract BRST level here and $x=a,\epsilon,v$ and
$\delta$ etc. was short for $s(x)$ as in Eq.\ (2.7) resp. (2.8).  As
discussed in the text, to obtain forms
$\omega_{2n-1-k,k}(a;X_1,\ldots,X_k)$ one has to evaluate the ghosts
$v$ at tangent vectors $X_j$ and, in the case of an odd Fredholm
module, take the trace in the auxiliary space with the Pauli matrix
$\sigma_3$. To complete the proof of Theorem 3.1 one needs to check
that the commutators mentioned in Theorem A5 --- which are on the
abstract BRST level --- turn into the commutators on the operator
level mentioned in Theorem 3.1. We now show that this is indeed the
case. For that we need to distinguish $s(x)$ from $x$. 
The commutators referred to in Theorem A5 are then of the
form $[s(\alpha),s(\beta)]$ where $s(\alpha)$ and $s(\beta)$ are
monomials in $s(v)$, $s(a)$ and $s(\epsilon)$.

In the case of an even Fredholm module, there is no extra auxiliary
space (cf. (2.7)), and the BRST commutators $[s(\alpha),s(\beta)]$ 
immediately become operator commutator after the 
evaluation of the ghosts at the tangent vectors.  

We thus can concentrate on odd Fredholm modules. Let $n_\alpha$ be
the ghost degree of $\alpha$ (the number of $s(v)$) and $d_\alpha$
the sum of the numbers of $s(\epsilon)$ and $s(a)$ (which is of course
only defined modulo $2$), and similarly for
$\beta$.  Then $s(\alpha)=\pm \alpha\otimes
\sigma_3^{d_\alpha}\sigma_1^{n_\alpha}$ and similarly for $\beta$, and
thus
$$
[s(\alpha),s(\beta)] = \pm \biggl((-1)^{n_\alpha d_\beta }\alpha\beta -
(-1)^{d_\alpha n_\beta } \beta\alpha \biggr)\otimes  
\sigma_3^{d_\alpha+d_\beta}\sigma_1^{n_\alpha+n_\beta}.
$$
In the odd case the ghost degree $k=n_{\alpha}+n_{\beta}$ is even
and $d_{\alpha}+d_{\beta}$ is odd. Evaluating $[s(\alpha),s(\beta)]$
for Lie algebra elements $X_j$, we obtain (up to an irrelevant overall
factor)
$$
\align 
\sum_{P\in S_k} {\text sign}_P &
\biggl(\alpha (X_{P(1)},\ldots, X_{P(n_\alpha)}) 
\beta (X_{P(n_\alpha+1)} ,\ldots, X_{P (k) })- \\
&(-1)^{n_\alpha n_\beta }(-1)^{n_\alpha d_\beta + d_\alpha n_\beta}
\beta(X_{P(n_\alpha+1)},\ldots, X_{P (k) })   
\alpha  (X_{P(1)},\ldots, X_{P(n_\alpha)}) \biggr)
\endalign 
$$
where $S_k$ is the permutation group of $k$ elements and ${\text sign}_P$ 
gives the parity of a permutation $P$. One now needs to show that
this always corresponds to a sum of commutators.
For that we need only to check that the exponent $n_{\alpha}n_{\beta}
+n_{\alpha}d_{\beta}+ n_{\beta}d_{\alpha}$ is always even. But this follows
from the fact that here $n_{\alpha}+n_{\beta}$ is even and $d_{\alpha}
+d_{\beta}$ is odd. 

\vskip 0.3in

REFERENCES 

\vskip 0.3in

[1] J. Schwinger: Field theory commutators. Phys. Rev. Lett. \bf 3,
\rm 296-297 (1959). 
S. Adler: Axial vertex in spinor electrodynamics. Phys. Rev. \bf 177,
\rm 2426 (1969). J. Bell and  R. Jackiw: A PCAC puzzle: $\pi_0\to
\gamma\gamma$ in the sigma model. Nuevo Cimento, \bf 60A, \rm 47
(1969). 

[2] J. Mickelsson: Chiral anomalies in even and odd
dimensions. Commun. Math. Phys. \bf 97, \rm 361 (1985).  L.D. Faddeev
and S. Shatashvili: Algebraic and Hamiltonian methods in the theory of
nonabelian anomalies. Theoret. Math. Phys. \bf 60, \rm 770 (1985) 

[3] K. Fujikawa: Path integral for gauge theories with
fermions. Phys. Rev. D (3), \bf 21, \rm no 10, 2848-2858
(1980). M.F. Atiyah and I. Singer:  Dirac operators coupled to vector
potentials. Proc. Natl. Acad. Sci. (USA) \bf 81, \rm no. 8, 2597-2600 (1984)

[4] J.M. Gracia-Bondia and  J.C.Varilly: On the ultraviolet behavior
of quantum fields over  noncommutative manifolds.  Internat. J. of
Modern. Phys. \bf A 14\rm, no. 8, 1305-1323 (1999) 

[5] A. Connes: \it Noncommutative Geomerty. \rm Academic Press, San Diego
(1994)

[6] E. Langmann: Descent equations of Yang-Mills anomalies in
noncommutative geometry. J. Geom. Phys. \bf 22, \rm no. 3, 259-279 (1997)

[7]  L. Bonora, M. Schnabl, and A. Tomasiello:  A note on consistent 
anomalies in noncommutative Y-M
theories. Phys. Lett. \bf B485, \rm 311-313 (2000).
J.M. Gracia-Bondia and C.P. Martin: Chiral gauge anomalies on
noncommutative $\Bbb R^4.$ Phys. Lett. \bf B479, \rm 321-328 (2000)

[8] J. Mickelsson: Wodzicki residue and anomalies of current
algebras. In: \it Integrable Models and Strings. \rm 
Springer Lecture Notes in Physics 436, 123-135, (1994) 

[9] E. Langmann and   J. Mickelsson: Elementary derivation of the
chiral anomaly. Lett. Math. Phys. \bf 36, \rm no. 1,  45-54 (1996) 

[10] A. Carey, M. Murray, and J. Mickelsson: Bundle gerbes applied to
field theory.  Revs. Math. Phys. \bf 12, \rm no. 1, 65-90 (2000)  

[11] H. Araki: Bogoliubov automorphisms and Fock representations of the
algebra of canonical anticommutation relations. Contemp. Math. \bf 62,
\rm 23-141, Amer. Math. Soc., Providence, RI (1987)

[12] B. Zumino: Chiral anomalies and differential
geometry. Relativity, Groups, and Topology, II (Les Houches 1983),
notes prepared by K. Sibold, 
1291-1322. North-Holland, Amsterdam and New York (1984)

[13] R. Palais: On the homotopy type of certain groups of
operators. Topology \bf 3, \rm 271-279 (1965)  

[14] J. Dixmier and A.  Douady: Champs continus d'espaces hilbertiens
et de $C^*$ algebres. Bull. Soc. Math. Fr. \bf 91, \rm 227-284  (1963) 

[15] A. Carey and  J. Mickelsson: A gerbe obstruction to quantization
of fermions on odd-dimensional manifolds with
boundary. Lett. Math. Phys. \bf 51, \rm 145-160 (2000)

[16] E. Langmann: Noncommutative integration
calculus. J. Math. Phys. \bf 36, \rm no. 7, 3822-3825 (1995)

[17] J. Mickelsson: Kac-Moody groups, topology of the Dirac
determinant bundle, and  fermionization. Commun. Math. Phys. \bf 110,
\rm no.2, 173-183 (1987)  

[18] M. Rieffel: Noncommutative tori - a case study of noncommutative 
differentiable manifolds. Contemp. Math. \bf 105, \rm
191-211. Amer. Math.
Soc., Providence, RI (1990) 

[19] M. Cederwall, G. Ferretti, B.E.W. Nilsson, A. Westerberg:
Schwinger terms and cohomology of pseudodifferential
operators. Commun. Math. Phys. \bf 175, \rm no.1, 203-220 (1996)

[20] S. Paycha: Renormalized traces as a looking glass into infinite
dimensional geometry. Preprint no. 671, Sonderforschungsbereich 256,
Bonn (2000).  A. Cardona, C. Ducourtioux, J.P. Magnot, and  S. Paycha: 
Weighted traces on the algebra of pseudodifferential
operators and geometry of loop groups.  math.OA/0001117  
 
[21] L.-E. Lundberg: Quasi-free ``second
quantization''. Commun. Math. Phys. \bf 50, \rm no.2, 103-112  (1976) 

[22] J. Mickelsson: \it Current Algebras and Groups. \rm Plenum Press,
London and New York (1989)

[23] R. Bhatia: \it Matrix Analysis. \rm Graduate Texts in Mathematics 
\bf 169, \rm Springer-Verlag, New York (1997)

[24]  J.M. Gracia-Bondia, J.C. Varilly, and  H. Figueroa: \it Elements of
Noncommutative Geometry. \rm Birkh\"auser, Boston (2001) 

\end

\enddocument